# Stimulated emission of radiation using spin-population inversion in metals: a spin-laser

V. Korenivski[1]*, A. Iovan[1], A. Kadigrobov[2]#, and R. I. Shekhter[2]
[1]Nanostructure Physics, Royal Institute of Technology, 10691 Stockholm, Sweden
[2]Condensed Matter Theory, Gothenburg University, 41296 Gothenburg, Sweden

Arrays of 10 nm-diameter point contacts of exchange-coupled spin-majority/spin-minority ferromagnetic metals, integrated into infrared-terahertz range photon resonators, are fabricated and measured electrically and optically. Giant, threshold-type electronic excitations under high-current pumping of the devices are observed as abrupt but reversible steps in device resistance, in many cases in access of 100%, which correlate with optical emission from the devices. The results are interpreted as due to stimulated spin-flip electron-photon relaxation in the system.

The discovery of masers [1,2] and lasers [3,4] has led to major breakthroughs in science and technology. Later important developments include compact semiconductor lasers for visible to near-infrared [5,6] and quantum-cascade lasers for far-infrared to THz radiation [7].

We have recently proposed a new principle of stimulated spin-flip photon emission in metallic ferromagnets, originating from the electron-photon interaction, with strength proportional to the large exchange energy in the ferromagnet, exceeding by orders of magnitude typical Zeeman energies for electrons in external magnetic fields [8]. A giant lasing effect is expected if the spin-population inversion is achieved by spin injection in a suitable nano-device, such as a photon resonator-integrated magnetic point-contact array [9]. The expected frequency range is dictated by the exchange splitting in the active ferromagnetic region, typically 10-100 THz, and the optical gain is expected to exceed that of conventional compact semiconductor lasers by 3 to 4 orders of magnitude due to the vastly higher carrier density in metals. If successful, such metal-based spin-flip lasers should be a breakthrough in the field of spin-photo-electronics.

We have recently reported experimental evidence for stimulated spin-flip photon emission in metals in the Zeeman-split configuration, with minority and majority spin injection into near-ballistic, ~10 nm-sized point contacts [10]. In this report we make the next step toward a spin-flip laser and integrate an array of magnetic point contacts into disk-shaped photon resonators we recently proposed [9], 10-50 μm in diameter. We observe giant, threshold-type resistance excitations for high current pumping correlated with infrared emission, consistent with the expected lasing action in the system.

Figure 1 illustrates the spin-flip photon emission process employed in this work. Electrons from a strong spin-majority ferromagnet (F) having the magnetic moments predominantly along the local magnetization, injected into a weak spin-minority ferromagnet (f*) having the conduction band polarized opposite to the magnetization of the lattice, create a spin-population inversion by populating the higher in energy minority band in f* (Fig. 1b), such as to allow transitions vertical in the electron momentum [8]. For this to occur the injected electrons must be of very high density and, importantly, non-equilibrium in energy. Therefore, the injection point contacts must be either ballistic or diffusive [10]. This in practice limits the contact size to approximately 10 nm, which is very challenging with regards to device fabrication. One of the spin-flip relaxation mechanisms of thus created inverse spin population is emission of photons, with energy corresponding to the exchange splitting in f*. This process is greatly enhanced if the emitted photons are preserved and localized using a suitable high-dielectric cavity for stimulating new spin-flip transitions [9]. The expected result then is strong stimulated spin-flip photon emission with a laser action.

The choice of the weak ferromagnet as spin-minority greatly simplifies the device design, in our opinion, as it removes the difficulty of achieving antiparallel alignment of the two magnetic electrodes in a nano-constriction. For F↑/f*↑, the two ferromagnets can be in direct contact, coupled in parallel by strong exchange across the interface, eliminating unwanted magnetization misalignments present in the alternative cases of F↑/N/f↓ or F↑/I/f↓ (N and I being a normal metal and insulator, respectively).

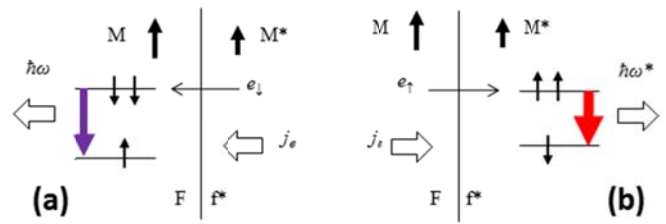

FIG. 1. (Color online) Schematic of spin-flip photon emission in a spin-majority/minority ferromagnetic point contact, for negative (a) and positive (b) bias. The arrows of different color and thickness illustrate the expected different frequency and intensity in the two cases.

Fig. 1a shows the configuration for the bias direction, in which the electrons are injected into the strong ferromagnet F. Due to the larger exchange splitting in F, the critical spin-population inversion is more difficult to achieve in this case, which therefore is expected to require a higher pumping current and lead to weaker spin-flip emission compared to the reversed case of spin-pumping into f* from F, shown in Fig. 1b. Thus, the spin-flip photon emission is expected to be bipolar in current polarity but have different threshold-bias conditions, intensities, and frequencies for the F/f* magnetic configuration chosen (different exchange spitting in F and f* dictate different emission frequencies).

We have developed a process for producing ~10 nm-range point contacts, embedded as an array into a dielectric matrix, integrated into photon resonators for the far-infrared to THz range radiation. It involves 4 lithography steps and the total of 70 process steps, described in details elsewhere [11]. In brief, the bottom and top thick metal electrodes are fabricated using photo lithography and the sensitive intermediate step for the active region uses colloidal lithography with a self-assembled monolayer of polystyrene particles. The resulting device is a 10-50 μm-diameter disk-shaped resonator with a composite active region of sub-10 nm magnetic point-contacts embedded into a dielectric matrix. The crossed bottom and top electrodes made of 200 nm thick Al are isolated electrically using a thin layer of silicon dioxide. The top electrode either fully or partially encloses the sides of the disk-shaped active region and thus serves as a light resonator, with strong gallery-whispering modes [9].

The active region is structured using a colloidal lithography mask, with the diameter of the individual openings in the Al-strengthened $SiO_2$ of approximately 10 nm, as shown by a SEM image in Fig. 2a (bottom panel). The openings serve as shadow masks for sputter deposition of the point contact material, as illustrated in Fig. 2b. We use $Fe/Fe_{70}Cr_{30}$ as the point contact core material, where Fe is the strong spin-majority ferromagnet (F in Fig. 1) and the $Fe_{70}Cr_{30}$ alloy is a weaker (lower magnetization/$T_C$) spin-minority ferromagnet (f* in Fig. 1), as previously reported by Voille et al. [12]. We estimate that the actual size of the nano-constrictions thus produced is somewhat smaller than 10 nm due to shadowing, which places the contacts in the spectroscopic range, ballistic-to-diffusive, with non-equilibrium in energy electron injection [13]. This size implies electrical resistance of approximately 10-20 Ω per point contact. A fully connected array of 10 μm diameter, with 2500 point contacts, would be expected to have resistance in the 10 mΩ range. We, however, purposely etch the colloidal mask to the extreme to achieve the small size needed for efficient non-equilibrium injection, such that only a fraction of the contacts are actually connected. The total array resistance is then typically of the order of 0.1 – 1 Ω (estimated 10-200 contacts in parallel), with the total injection current of up to ~100 mA, and estimated up to ~10 mA (~100 mV) per contact.

Such a composite metal/dielectric active region has greatly enhanced properties compared to conventional lasers [9], with two key advantages being that the enhanced photon field of the high-dielectric, highly-transparent resonator is strongly coupled to the point-contact cores, while the heat is efficiently transferred by electron and phonon drain into the massive metallic electrodes.

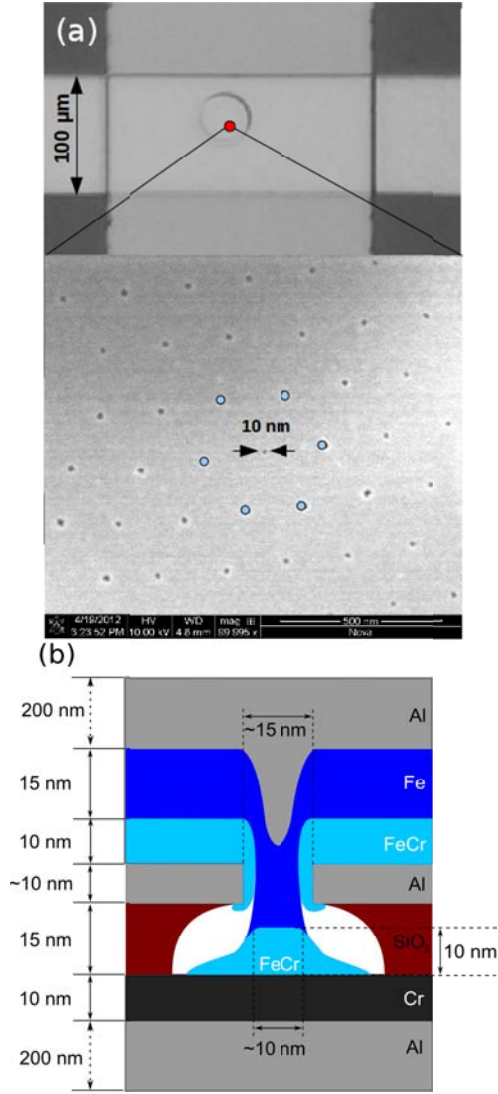

FIG. 2. (Color online) (a) Optical image of an integrated closed-cavity disk-shaped resonator, with the thick bottom and top Al metal electrodes acting as full-reflection mirrors. The zoom-in shows a section of the active region mask, with a hexagonal-close-packed array of point-contact openings, ~10 nm-diameter each, with 200 nm inter-contact separation. The patterned $SiO_2$ layer, reinforced with a thin Al layer, serves as the shadow mask for deposition of the point contact core. (b) Layout of a single point contact, with the magnetic contacts acting as the photon source and the embedding dielectric as the propagation/resonator medium. The structure in the form of a 10-50 μm disk-shaped point-contact array is then enclosed by the thick metal top and bottom electrodes, shown in (a).

For the above design, with strong enough spin injection and high non-equilibrium spin accumulation in the contact core, any photon emitted by a spin-flip process is contained within the device and enhanced by the high dielectric constant of the oxide matrix of the active region. This photon, due to the high transparency of the oxide matrix, is long lived and therefore has a high probability to stimulate another spin-flip event. At a critical injection current, the lifetime of the emitted photon exceeds the effective photon absorption rate in the resonator and a cascading avalanche of spin-flip transitions can then take place, generating high-density electromagnetic modes in the cavity. This threshold for the stimulated spin-flip photon emission is expected to manifest itself as threshold-type changes in the physical characteristics of the device, such as its current-voltage characteristics. We indeed observe such threshold-type excitations, correlated with optical emission from the devices, as detailed below.

Figure 3a shows the differential resistance of a 36 μm-diameter resonator. The zero-bias resistance is ~1 Ω and increases monotonously with bias, characteristic of phonon relaxation (heating) in point contacts [13]. The current densities are very high when compared to the typical current carrying capacity in bulk metals, up to ~$10^{10}$ A/cm$^2$ per point contact in this case. This exceeds the melting current densities in the respective bulk metals and is possible in point contacts due to the good electronic and thermal contact with the massive metal electrodes, serving as vast electron and phonon baths. The above extreme current density is the key in achieving the critical spin-population inversion in the system [8,10]. When that occurs, photon emission and stimulated photon emission can take place, accompanied by an abrupt increase in the radiation power density in the resonator. At this point the rate of stimulated emission can exceed the photon absorption in the resonator, resulting in a lasing action in the device, visible as abrupt step-like changes in the differential resistance as well as directly in the current-voltage characteristics (Fig. 3b). The additional power transferred into the photon field from the electrical injection increases the effective resistance of the device and, estimating from the I-V, can reach the order of 10 mW.

The magnitude of the resistance changes observed are giant compared to any know relaxation processes in metallic point contacts [13], which we interpret as follows. In the enclosed resonator the photon field builds up with the only dissipation channel being generation of heat in the skin depth of the metallic mirrors/electrodes. (Stimulated emission dominates the photonic processes in the device, but itself leads to no dissipation of photons.) At the threshold injection the lasing photon field eventually dissipates into heat, raising the temperature of the device. A rise in temperature of 100-200 degrees can be expected in a 10-100 μm size device dissipating ~10 mW, and would naturally raise the temperature of the point contacts. This in turn would lead to an *abrupt* ~2-fold increase in the contact

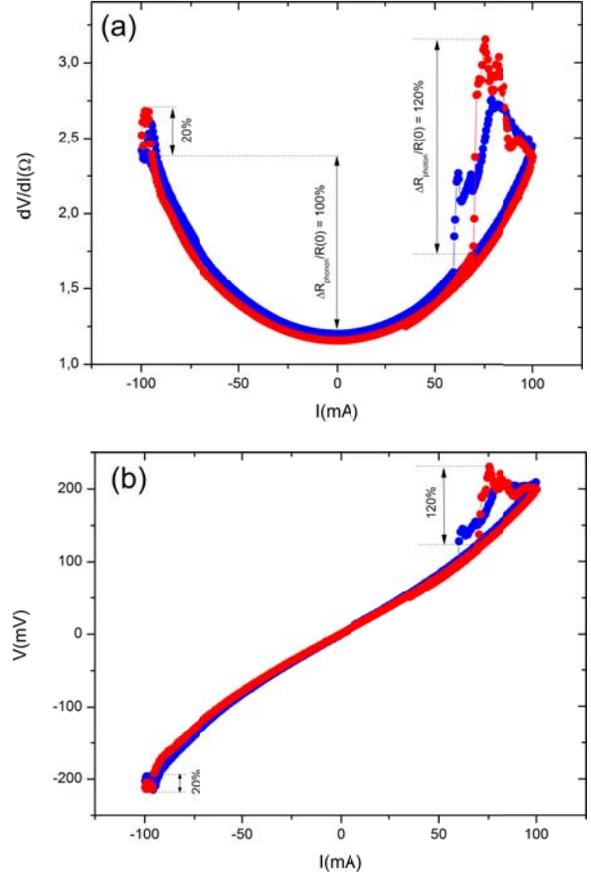

FIG. 3. (Color online) Differential resistance (a) and current-voltage characteristics (b) of a 36 μm-diameter resonator, recorded in repeated sweeps, on two successive days (red, blue).

resistance, comparable to the *smooth* 2-fold phonon heating seen at maximum negative bias.

An interesting feature of the observed I-V characteristics is the fall off in the resistance at still higher bias and no excitations on the return path to low bias. This is consistent with the above interpretation, since the heated point-contact core has reduced or fully diminished magnetization due to thermal disorder, and would not therefore possess the spin-flip photon resonance providing the original emission. We note that the relevant spin-splitting in f* (F) is proportional to a product of the exchange constant and the magnetization, which is a function of temperature [10]. After cooling off at low bias, repeated I-V sweeps reproduce the effect, with some devices successfully tested over hours and days, as shown in Fig. 3. The data also show that the effect is bi-polar, as expected (see Fig. 1), with the threshold bias and intensity different for different polarities.

A number of resonators of varying size, measured at room and liquid nitrogen temperature, showed pronounced resistance excitations (~100%). Thus, the effect was found to be relatively insensitive to the ambient temperature, which is to be expected since strongly current-pumped nano-contacts can develop temperatures tens and hundreds of degrees higher than the cryostat temperature [13]. The devices are observed to "tune in" into the spin-flip resonance on moderate heating (lowering of the effective exchange splitting by thermal phonons) and "tune out" out of the resonance on further (photon) heating. All this takes place above the room temperature, toward the Curie point of the weak ferromagnet (f*, FeCr).

We do not observe the effect at RT in our smallest resonators, ~10 μm in diameter, which suggests the lower boundary for the excited photon wavelengths. The fact that a range of the resonator sizes, 20-50 μm diameter, exhibit the effect indicates that either the spin-flip transition is not very sharp, possibly due to a contact-to-contact variation in properties, or multiple (gallery whispering) modes are excited in the cavity as is in fact expected [9].

We have fabricated calibration samples with the identical layout but with the point contact core replaced by a nonmagnetic material (Cu) and verified that they do not show any excitations of the type discussed above, but only the expected near-parabolic phonon heating. This indicates that the strong excitations observed are magnetic in nature.

In order to directly verify the optical nature of the observed effect we have fabricated and measured partially open resonators, illustrated in Fig. 4a. Compared to the fully enclosed resonators of Fig. 2, here the top electrode is displaced by a fraction of the diameter and leaves the $SiO_2$ matrix open to radiate. The electro-magnetic modes in the composite matrix, visualized using a numerical simulation of the device (see also [9]), should have a high intensity at the opening. A pyroelectric sensor (two-element differential, no optical filer) is faced to the opening for detecting radiation from the device, in the far-infrared to THz range (>5 μm).

Fig. 4b shows the pyro-sensor signal, measured with a 4 Hz chopper between the sample and the sensor and the lock-in time constant of 1 second. This arrangement is sensitive to both the relatively slow-varying IR-radiation due to the heating of the sample as a whole and direct spin-flip radiation from the resonator. The former is parabolic and has the typical time constant of ~100 sec (not shown; often nearly identical to the 'phonon-parabola' in Fig. 3a). The latter is due to electronic transitions and should have orders of magnitude sharper onsets. The data shows a peak in the IR signal, correlated with the 50% excitation in the resonator's electrical resistance, on top of the nearly constant thermal-background signal of ~1.1 mV.

A 42 μm diameter resonator, measured differentially, without the chopper, sensitive only to changes in the IR-radiation on the scale of ~1 Hz or faster, shows large spikes in the IR-signal, correlated with the large-magnitude conduction excitation. This detection method subtracts the slowly-varying thermal component from the IR-signal, sensing the fast component, which must be of electronic origin as it correlates well with the conduction excitations. The measured optical data support the above interpretation of stimulated spin-pumped photon emission in the device.

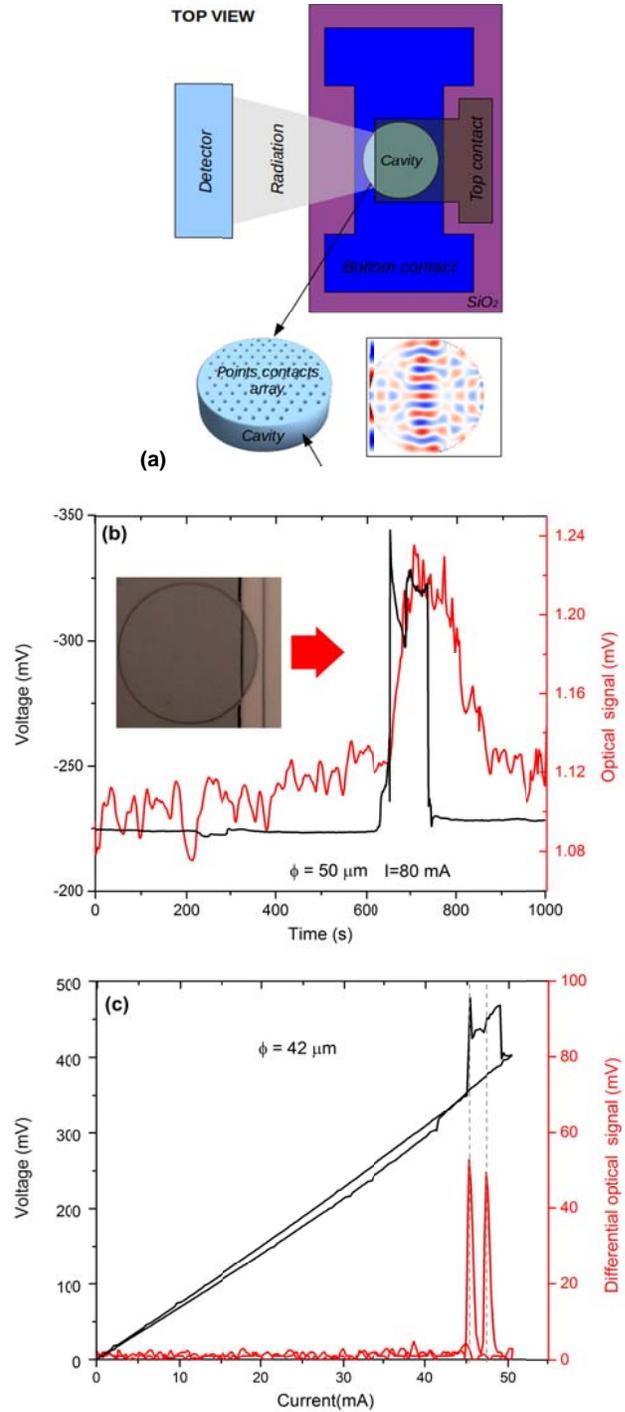

FIG. 4. (Color online) (a) Schematic of a partially open resonator, including a visualization of the electromagnetic modes' intensity in the matrix/cavity (red and blue are high and low intensity, respectively [14]). A number of modes are excited, with a high intensity expected at the opening. (b) Voltage across a 50 μm diameter resonator at constant pumping current of 80 mA, showing an excitation event, with a simultaneous IR-detector voltage. Inset is a microscope image of the resonator. (c) V-I characteristic for a 42 μm diameter resonator, with a simultaneous differential IR detection.

In conclusion, we have fabricated ballistic-range magnetic point contact arrays integrated in IR-THz range metal-oxide resonators and observed threshold-type I-V excitations of giant magnitude, correlated with optical emission from the devices, consistent with the expected spin-flip laser effect in the system.

We gratefully acknowledge financial support from EU-FP7-FET-Open through project Spin-Thermo-Electronics.

*Corresponding author: vk@kth.se.
#Also with: Teoretische Physik III, Ruhr-Universitaet, D-44780 Bochum, Germany.